\documentclass[multphys,vecphys]{svmult}

\usepackage{makeidx}         
\usepackage{graphicx}        
\usepackage{multicol}        
\usepackage[bottom]{footmisc}

\makeindex             

\begin{document}

\title*{Hydrodynamical model atmospheres\\and 3D spectral synthesis}
\author{Hans-G{\"u}nter Ludwig\inst{1} \and\ Matthias Steffen\inst{2}}
\institute{CIFIST, GEPI, Observatoire de Paris-Meudon, 92195 Meudon Cedex, France
\texttt{Hans.Ludwig@obspm.fr}
\and 
Astrophysikalisches Institut Potsdam, 14482 Potsdam, Germany
\texttt{msteffen@aip.de}}
\maketitle

\section{Radiation-hydrodynamics modeling -- overview}
\label{s:overview}

In this paper we discuss three issues in the context of three-dimensional (3D)
hydrodynamical model atmospheres\index{hydrodynamical model atmosphere} for
late-type stars, related to spectral line shifts, radiative transfer in
metal-poor 3D models, and the solar oxygen abundance. To establish the context
we start by giving a brief overview about the model construction, taking the
radiation-hydrodynamics code \mbox{CO$^5$BOLD}\ (COnservative COde for the
COmputation of COmpressible COnvection in a BOx of L Dimensions with
\mbox{L=2,3}; \cite{FSD02}) and the related spectral synthesis package
Linfor3D as examples.

Based on a Godunov-type finite volume approach, \mbox{CO$^5$BOLD}\ provides the
time-dependent solution for a one-component compressible radiating fluid in an
external gravity field on a fixed, non-staggered 3D Cartesian grid (allowing
variable spacing). Operator splitting separates Eulerian hydrodynamics,
optional tensor viscosity, and radiation transport.  Directional splitting
decomposes the 3D hydrodynamics problem into 1D sub-steps which are treated by
an approximate Riemann solver of Roe type, modified to work with an arbitrary
equation of state, and to properly handle an external gravity field. This
scheme is very robust and well adapted to handle transonic flows and shocks in
a highly stratified medium. By design, the code guarantees the numerical
conservation of mass, momentum, and energy.  For any prescribed chemical
composition, \mbox{CO$^5$BOLD}\ uses a tabulated equation of state taking into account
partial ionization of H\,I, He\,I, and He\,II, as well as the formation and
dissociation of H$_2$ molecules. 

The role of radiation in the hydrodynamical simulations is to describe the
energy balance due to radiative heating and cooling. The radiative energy
exchange is computed from the solution of the non-local transfer equation on a
system of a large number of rays traversing the computational volume under
different azimuthal and polar angles. Realistic stellar opacities are used,
optionally based on ATLAS or MARCS opacity data. The frequency dependence of
the radiation field is treated in a multi-group approximation -- the so-called
\textit{opacity binning method\index{opacity binning method}} (OBM;
\cite{N82,LJS94,VBS04}) -- where frequencies are sorted into a small number of
bins (typically 4 \ldots 6) according to the ratio of monochromatic to
Rosseland optical depth. So far, strict LTE is assumed, thus scattering cannot
be treated. Radiation pressure is ignored.

The code Linfor3D accepts \mbox{CO$^5$BOLD}\ models as background structures
on which spectral synthesis calculations at high wavelength resolution --
usually focusing on one particular spectral line -- can be performed. When
calculating the emergent spectrum, Linfor3D takes into consideration the full
3D flow geometry including Doppler shifts caused by the macroscopic
hydrodynamical velocities.  It represents the effects of thermal and pressure
broadening in standard fashion, but leaves out the ad-hoc broadening
mechanisms of micro- and macro-turbulence introduced in 1D atmospheric models.
Similar to \mbox{CO$^5$BOLD},strict LTE is assumed in Linfor3D. Resulting
spectral line profiles provide detailed information about intrinsic line
shapes, and convective line shifts with respect to a line's laboratory
wavelength.

\vspace{-0.5\baselineskip}

\section{High precision line shifts from 3D models?}
\label{s:lineshifts}

\begin{figure}[t!]
\centering
\includegraphics[width=0.45\textwidth]{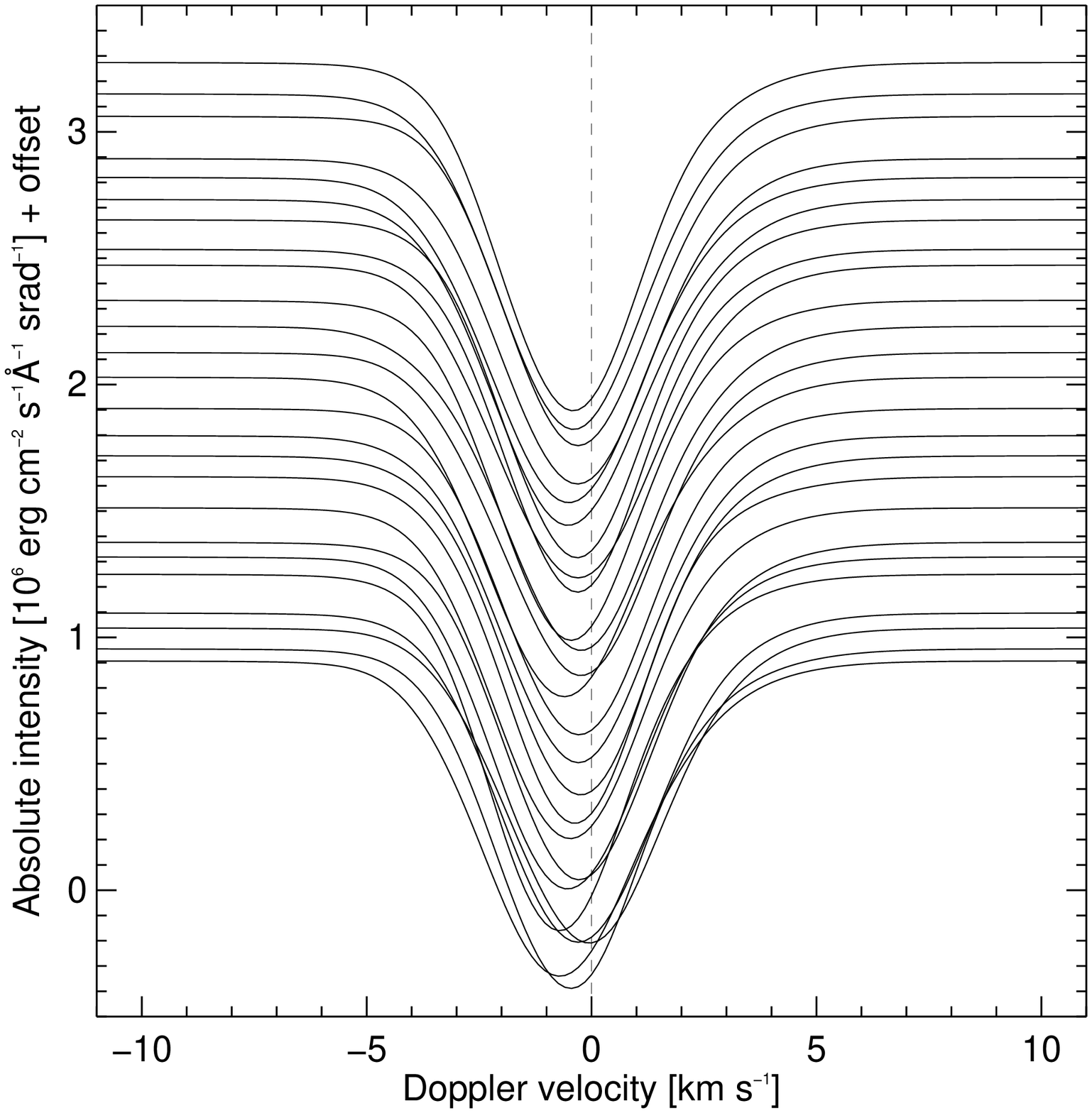}\hspace{4ex}%
\includegraphics[width=0.45\textwidth]{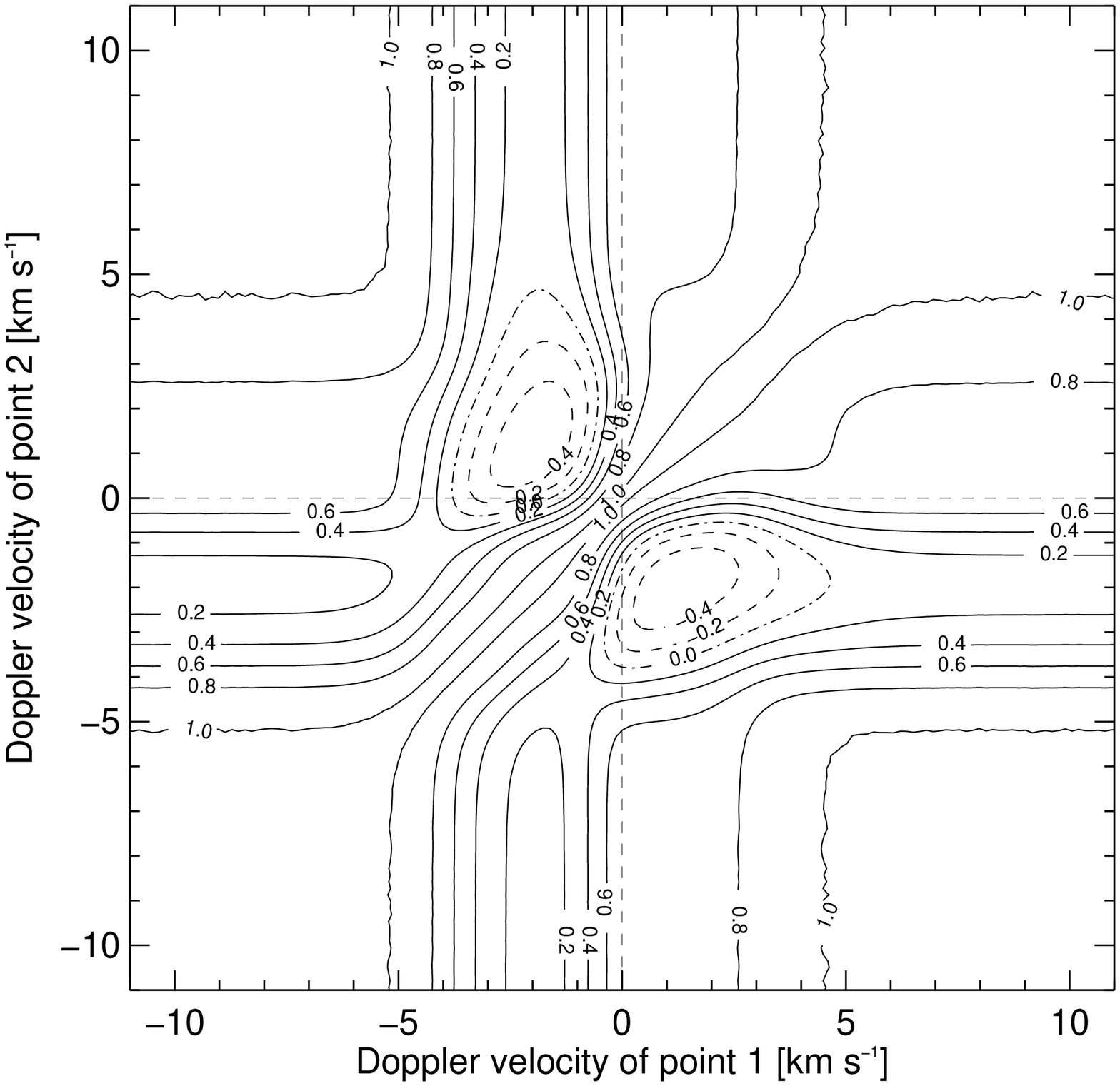}%
\caption{\textbf{Left panel:} Time series of line profiles of
  a Fe\,I line at 6082\,{\AA} in a 3D solar \mbox{CO$^5$BOLD}\ model. The wavelength is given as
  corresponding Doppler velocity with respect to the line's laboratory
  wavelength. The line profiles have been offset proportional to time running
  from top to bottom. The time interval between profiles is the same,
  fluctuations in the continuum brightness cause the non-equidistant
  appearance. \textbf{Right panel:} Contour plot of correlation coefficients of the
  intensity between two wavelength points in the profiles shown in the left panel.}
\label{f:lineshifts}
\end{figure}

A \mbox{CO$^5$BOLD} simulation constitutes a statistical realization of the
atmospheric flow field in the stellar surface layers. If one is not interested
in studying time-variable phenomena but only in the mean state of the
atmosphere, the fluctuations present a noise source which limits the precision
to which flow and related spectroscopic properties can be determined. This is
similar to the observational situation where the intrinsic variability of a
star limits the precision to which its radial velocity, e.g.  in planet
searches, can be measured.  The left panel of Fig.~\ref{f:lineshifts}
illustrates the temporal variability of a Fe\,I line calculated from a
hydrodynamical solar model. Each line profile is a horizontal average over the
surface of the computational box.  Shown are 25 instants in time which are
sufficiently separated that they can be considered statistically uncorrelated.
We ask: what is the precision due to the statistics (ignoring systematic
effects) to which we can determine the line shift\index{line shift}?

It is straight forward to show (see \cite{L06}) that the expectation value of
the disk-integrated line profile corresponds to the expectation value of the
profile of the local hydrodynamical model. Hence, the statistical
uncertainties of the profile emerging from the model directly correspond to
the uncertainties of the predicted disk-integrated profile. From
Fig.~\ref{f:lineshifts} it is obvious that the the statistical fluctuations
are not just pixel-to-pixel random noise like in the case of photometric
Poisson noise. The line profiles change their overall shape, i.e. different
wavelength points show a considerable degree of correlation. The linear
correlation coefficient between intensities at two wavelength points 1 and 2
is given by
\begin{equation}
C\left[I_1,I_2\right]\equiv
\frac{\langle\Delta I_i\,\Delta I_j\rangle}{\sigma_{I_1}\sigma_{I_2}}
= \frac{\langle I_1 I_2 \rangle
  -\langle I_1 \rangle \langle I_2 \rangle}{\sigma_{I_1}\sigma_{I_2}}.
\label{e:cordef}
\end{equation}
$I$ denotes the intensity, $\langle .\rangle$ the temporal average. $\Delta I_i \equiv I_i-\langle I_i
\rangle$ is the intensity deviation from the mean. The right
panel of Fig.~\ref{f:lineshifts} shows the correlation matrix of the
example line depicted. 

In order to quantify the line shift~$\lambda_\mathrm{s}$ we need a model of
the procedure by which it is measured, which in turn emerges from the
definition of $\lambda_\mathrm{s}$. Here, we assume that the measuring
procedure of $\lambda_\mathrm{s}$ can be described by a function $\Lambda$ of
potentially all available (assumed discrete) intensities $I_i$:
$\lambda_\mathrm{s}=\Lambda\left(I_i\right)$. In order to make algebraic
headway we simplify and linearize $\Lambda$ around the expectation value of
the line profile described by the values $\langle I_i \rangle$. To leading
order in $\Delta I$ we obtain for the variance of the line
shift the standard expression of the error propagation for correlated
variables
\begin{equation}
\sigma^2_{\lambda_\mathrm{s}}\approx \sum_{i,j}\frac{\partial\Lambda}{\partial
  I_i} \frac{\partial\Lambda}{\partial I_j}\,\langle\Delta I_i\,\Delta I_j\rangle
= \sum_{i,j}\frac{\partial\Lambda}{\partial
  I_i}\frac{\partial\Lambda}{\partial I_j}
\,\sigma_{I_i}\sigma_{I_j}\,C\left[I_i,I_j\right].
\label{e:varshift}
\end{equation}
The summation is performed over all pixels which are relevant for the
measurement of $\lambda_\mathrm{s}$.  Equation~(\ref{e:varshift}) emphasizes
the role of the covariance matrix of the intensities $\langle\Delta
I_i\,\Delta I_j\rangle$ -- or equivalently the standard deviations of the
intensities and their correlation matrix -- plays for the magnitude of the
uncertainty of the line shift. In the present context we discussed line shifts
but relation~(\ref{e:varshift}) of course also holds for other measures like,
e.g., the equivalent width of a line. The statistical quantities in
relation~(\ref{e:varshift}) can be estimated from the time series provided by
the hydrodynamical model.  Asymptotically, for a given line one will arrive at
a fixed value for the correlation matrix $C\left[I_i,I_j\right]$. If one
wants to improve the accuracy of the line shift one has to beat down the
uncertainties in the intensities $\sigma_{I_i}$. This can be achieved by
longer time series or larger horizontal extent of the hydrodynamical model.

Our example Fe\,I line shows a RMS temporal scatter of its position of
0.16\,km\,s$^{-1}$. The value was obtained by directly (and somewhat
heuristically) measuring the location of the line core without formalizing the
process by explicitely constructing a measurement function~$\Lambda$. The
statistical independence of the 25 individual snapshots implies an uncertainty
of about 30\,m\,s$^{-1}$ for the line shift. While the specific value depends
on the chosen line and selection of snapshots we think that it gives an
indication of the precision one is typically working with in todays
hydrodynamical standard models. Higher precision is possible but
computationally also more costly. Of course, at some point real uncertainties
will be dominated by systematic shortcomings of a model.

\vspace{-0.5\baselineskip}

\section{3D radiative transfer in metal-poor atmospheres}
\label{s:obm}

As mentioned earlier, the radiative transfer in the 3D models is commonly
approximated by the opacity binning method (OBM) assuming strict LTE. While
the approach is working fine in atmospheres of about solar metallicity,
metal-poor atmospheres pose a challenge to the OBM. At first glance, this may
come as surprise because the dramatic decrease of the number of spectral lines
relevant for the radiative energy exchange should simplify the radiative
transfer. However, the actual situation is quite different. First, scattering
in the continuum becomes important for the thermal structure of metal-poor
atmospheres\index{metal-poor atmospheres}. In the OBM, scattering is treated
as true absorption so that one must expect some effects on the resulting
temperature structure. Second, experience has shown that the OBM does not work
as accurately in metal-poor atmospheres as in atmospheres of solar
metallicity. It turned out that this deficit is not related to the treatment
of the line blocking but already shows up for the radiative transport in the
continuum.

\begin{figure}[t!]
\centering
\includegraphics[width=0.7\textwidth]{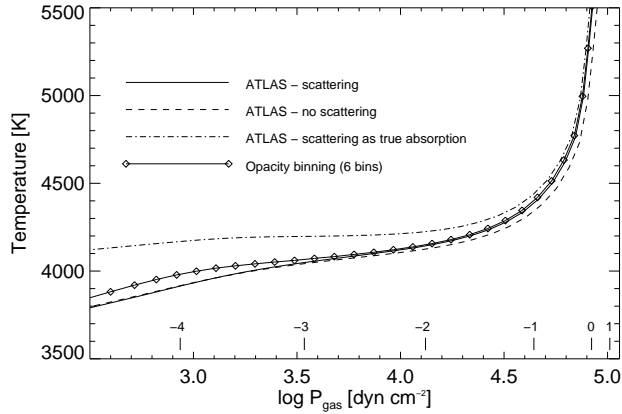}
\caption{Comparison of 1D model atmospheres ($T_\mathrm{eff}=5000\,\mathrm{K}$, $\log(g)=2.94$,
$\left[\mathrm{M/H}\right]=-2$) in radiative-convective equilibrium based on
different treatments of the radiative transfer. (for details see text)}
\label{f:scattering}
\end{figure}

Figure~\ref{f:scattering} shows an example of an atmosphere of a metal-poor
giant. Plotted are temperature profiles of 1D ATLAS6 (see~\cite{K79}) model
atmospheres in radiative-convective equilibrium. The only difference among the
models is the way in which the radiative transfer was treated. In three cases
labeled ``scattering'', ``no~scattering'', ``scattering as true absorption'' a
high wavelength resolution was employed, and scattering was treated exactly,
scattering opacity was neglected, or treated as true absorption, respectively.
The by far dominating scattering opacity under the considered conditions is
Rayleigh scattering by hydrogen atoms.  As evident from
Fig.~\ref{f:scattering}, the temperature structure is noticeably
influenced by scattering. The OBM used in the 3D models was also implemented
in the 1D atmosphere code and a resulting radiative-convective equilibrium
calculated. Comparison with the exact radiative transfer solution shows a
close correspondence from the deep layers up to lower optical depth of
$\log\tau\approx -3$. However, while useful in practice this is only
fortuitous. The OBM based temperature structure should in fact follow the
``scattering as true absorption'' case since in the OBM scattering opacity is
treated as true absorption. At present, the reason for the rather poor
performance of the OBM is unclear. Identifying its cause, improving the OBM,
and including scattering effects in 3D geometry are challenges to be met in
near-future developments of 3D model atmospheres for metal-poor stars.

\vspace{-0.5\baselineskip}

\section{3D models and the solar oxygen abundance}
\label{s:oxygen}

Recent spectroscopic abundance determinations by \cite{AGS04}, based on a 3D
hydrodynamic model atmosphere, led to a much debated downward revision of the
solar C, N, and O abundances. Their result for the oxygen
abundance\index{solar oxygen abundance} is $\log \epsilon_{\rm
  O}$\,=\,8.66\,$\pm$\,0.05 (on the scale $\log \epsilon_{\rm H}$\,=\,12),
causing a dramatic deterioration of the agreement between the thermal
structure derived from helioseismic inversions and theoretical solar models,
respectively.  Motivated by this problem, we (see~\cite{CLS07}) are currently
using a 3D \mbox{CO$^5$BOLD} simulation with 5-bin frequency-dependent
radiative transfer based on MARCS opacities to see whether the results by
\cite{AGS04} can be confirmed.  This independent redetermination of the solar
oxygen abundance is based on 2 forbidden and 7 permitted O\,I lines, using a
number of different observations, including both disk-center (``intensity'')
and full-disk (``flux'') spectra. In addition to 25 snapshots from the
simulation, we also derive abundances from different 1D atmospheres for
comparison.  Special care is taken to provide realistic error estimates.

The following preliminary conclusions can be drawn at this point: (i)
``intensity'' and ``flux'' spectra give practically the same result.  (ii) the
oxygen abundance derived from the 3D CO$^5$BOLD simulation is only slightly
lower (by --0.04\,dex) than that derived from the 1D empirical model by
\cite{HM74} (hereafter HM), indicating that the 3D mean model and the 1D HM
model have very similar temperature structures in the relevant layers. (iii)
the unknown cross sections for neutral particle collisions introduce
uncertainties in the NLTE corrections for the O\,I triplet lines of up to 0.1
dex; depending on the weight of these lines, the resulting error in the mean
oxygen abundance derived from our set of lines is about 0.05 dex.  (iv) Our
preliminary best estimate for the solar oxygen abundance is $\mathbf{\log
  \epsilon_\mathrm{O}=8.72\pm 0.06}$, which is close to the value recommended
by \cite{H01}, $\log \epsilon_\mathrm{O}=8.736\pm 0.078$. A remaining problem
of our analysis is that the two forbidden $[$O\,I$]$ lines give significantly
different abundances, which cannot be explained by NLTE-effects or
deficiencies of the model atmosphere. We hope to resolve this problem by
analyzing the observed center-to-limb variations of these two line profiles.

\vspace{-0.5\baselineskip}

\section{Remarks on precision spectroscopy and 3D models}

Hydrodynamical model atmospheres are on their way of becoming a standard tool
for the analysis of stellar spectra. Their ability of making detailed
predictions about the shape of spectral lines in convective atmospheres can
only fully exploited if observed spectra of sufficient resolution are
available. Ideally, spectrographs should be able to provide a spectral
resolutions above $10^5$ -- something that we would like instrument builders
to keep in mind.

High-fidelity abundance work benefits from the theoretical knowledge of the
precise intrinsic line shape. However, in practice one is nonetheless often
confronted with ambiguities, e.g. in the case of blends, which remain
unresolved by considering disk-integrated line profiles only.  The
center-to-limb variation of a line shape can provide crucial further
constraints. Combining interferometry with high-resolution spectroscopy (like
in the UVES-I project presented by A.~Quirrenbach, this volume) can open-up
this source of information for stellar work.

\vspace{-0.6\baselineskip}

\printindex
\end{document}